\documentclass[12pt, preprint]{aastex}

\newcommand{\HII}{H\,{\sc ii}}
\newcommand{\HI}{H\,{\sc i}}

\begin{document}

\title{Brackett Lines from the Super Star Cluster Nebulae in He 2-10}

\author{ Alaina L. Henry\altaffilmark{1}, Jean L. Turner \altaffilmark{1},
Sara C. Beck\altaffilmark{2}, 
Lucian P. Crosthwaite\altaffilmark{3}, 
\& David S. Meier \altaffilmark{4,5}}

\altaffiltext{1} {Department of Physics and Astronomy, Box 951547, UCLA, Los Angeles, CA 90095;
 ahenry@astro.ucla.edu, turner@astro.ucla.edu}
\altaffiltext{2}{Department of Physics and Astronomy, Tel Aviv University, 69978 Ramat Aviv, Israel; sara@wise1.tau.ac.il}
\altaffiltext{3}{Northrop-Grumman, San Diego, CA; lpcrosthwaite@cox.net} 
\altaffiltext{4} {National Radio Astronomy Observatory, P.O. Box 0, Socorro, NM 87801; dmeier@nrao.edu}
\altaffiltext{5}{Jansky Fellow}

\begin{abstract}
We present high spectral resolution ($v \sim 12-16~ {\rm km~s}^{-1}$) Brackett line
 spectroscopy of the blue compact dwarf galaxy Henize 2-10 made with NIRSPEC
 on the Keck Telescope.  
 The spatial resolution is seeing limited, at 1\arcsec.
 We detect two distinct kinematic features separated by approximately
3\arcsec, with heliocentric velocities of $\sim$860 and $\sim$890
km s$^{-1}$.  In addition to a narrow core, the line profiles also display a broad, 
low intensity feature on the blue side of the centroid,
which we attribute to an outflow.   This may be a sign of aging
in the clusters.     We compare to archival high resolution Very Large
Array (VLA) data at 1.3 cm,  and find that the centimeter wavelength emission is resolved
into six sources.  These radio sources are organized into two
 larger groups, which we associate with the two kinematic peaks in the Brackett spectrum.    
 We estimate  a Lyman continuum rate of  at least $7\times 10^{52} ~{\rm s}^{-1}$,  with a corresponding stellar 
 mass of $6 \times 10^6~M_{\sun}$ is required to ionize the nebulae.   We also estimate the size of 
 the nebulae from the radio continuum brightness  and find that the observed sources
  probably contain many  \HII\  regions in smaller, unresolved clumps.   Brackett line profiles have
   supersonic line widths, but, aside from the blue wing, are comparable to line widths observed
   in Galactic ultracompact \HII\ regions, which are excited by a single star, or a few stars.       
 \end{abstract}


\section{Introduction}
Starbursts are important events in galaxy evolution; they can
contribute significantly
to the total bolometric luminosities of even large galaxies. The
physical effects of concentrations of
massive stars on galaxies and on future generations of
star formation are not well understood. Stellar winds and supernovae
allow chemical enrichment of the gas, and
dissociation, ionization, and chemical processing of molecular clouds;
this feedback
can serve either to disperse the clouds and arrest the star formation,
or to compress
the clouds and promote new star formation.
Local starburst galaxies are excellent laboratories in which to study
the causes of starbursts
and their feedback, since at subarcsecond resolution, individual \HII\
regions, molecular clouds,
star clusters, and OB associations can be resolved.

He 2-10 is a  nearby (9 Mpc) blue compact dwarf galaxy undergoing a massive starburst.  
Optical imaging shows two regions of star formation, separated by
 approximately 10-12\arcsec \citep{Corbin}.  The western region (A)  displays signs of a 
 recent star formation episode.    
The presence of Wolf-Rayet (WR) stars \citep{Allen76} and thermal radio
 emission \citep{JK03} indicate the extreme youth of the burst, and H$\alpha$ imaging
  reveals large scale bubbles \citep{Mendez99}.

 Several young super star clusters within region A have been found in optical and 
 UV studies 
(\citealt{Chandar}; \citealt{Johnson00}; \citealt{VC92}), but these observations pick out objects 
which are relatively unobscured, and are probably not the youngest star-forming regions.   
In an effort to penetrate the obscuring material, 
infrared and radio observations have been made (\citealt{Beck01}; \citealt{Davies}; \citealt{JK03}), 
 revealing 
several radio/infrared nebulae.   These \HII\ regions are organized in
 two complexes separated by $\sim3-4\arcsec$, and can
be subdivided into several clumps.  Mid-infrared L\arcmin\ observations
 by \cite{Cabanac} successfully bootstrap the
  precise radio positions to the infrared and optical, and show that some of
   the radio sources may be without near infrared or optical counterparts.  
   However,  with a positional uncertainty of $\sim$ 15 pc, super star-clusters, 
   which are expected to be as small as 1 pc, are difficult to identify with infrared
    sources.    
In addition, single dish radio observations show a nonthermal spectral index,
 indicating the presence of SN remnants, and suggesting significant amounts of prior 
 star formation \citep{Allen76}. 
 He 2-10 also displays  a streamer of infalling molecular gas suggestive of a past
  merger (\citealt{K95}; \citealt{Meier}).  On the other hand, the \HI\ gas is organized in a ring,  
  and a condensation of CO may be filling the central hole \citep{K95}, much like in normal spiral
   galaxies (e.g., \citealt{ML78}).

In this paper, we use high spectral resolution observations
of He 2-10 at Br$\gamma$ and Br$\alpha$, made with 
NIRSPEC on Keck II,  in combination with an archival 1.3 cm Very Large Array (VLA) map  to constrain 
the masses, sizes, and densities of several compact \HII\ regions found in the dwarf starburst 
galaxy He 2-10.  These cluster properties, combined with the Brackett line profiles 
provide a better understanding of the  kinematics of the starburst, and can
 give clues to the history and evolution of young clusters.
   In \S \ref{obs} we summarize our NIRSPEC  observations and calibration 
   of the archival VLA data.  In \S \ref{results} we present our results, 
   and in \S \ref{discussion} a discussion of the energetics and kinematics.  
   We adopt a distance of 9 Mpc, which gives a scale of 44 pc/\arcsec.  
    Heliocentric velocities are used throughout.

\section{Observations \& Data Reduction}
\label{obs}
\subsection{NIRSPEC Observations \& Data Reduction}
\label{nirspecobs}
The infrared data were taken with the NIRSPEC, a 
cross-dispersed cryogenic echelle spectrometer on Keck II (\citealt{McLean98}, 2000).  
The Br$\gamma$ spectra were taken in the NIRSPEC-6 filter (1.56 - 2.30 \micron) on 
the night of December 12, 2000, and the Br$\alpha$ spectra were taken on the night 
of February 10, 2003, using the KL filter (2.16 - 4.19 \micron).   Observations 
were made in high resolution mode, with slit widths  0\arcsec.6, and 0\arcsec.4 , and position 
angles of 80 and 100 degrees east of north for Br$\alpha$ and Br$\gamma$, respectively.  
Spectra were taken in steps across He 2-10, with nine positions in Br$\alpha$ and three
 positions in Br$\gamma$.  The spectra were nodded, with the target both on and off of 
 the slit, to better enable the removal of background emission and sky features.  The spectral
  resolution is 12  km s$^{-1}$ for the Br$\gamma$ observations and 16 km s$^{-1}$ for Br$\alpha$. 
    For both sets of observations,  the airmass was 1.6.

The NIRSPEC slit viewing camera (SCAM) provided simultaneous infrared 
imaging of He 2-10, including the locations of the slit.   These broadband infrared
 images do not resolve the \HII\ regions seen in the radio, but rather, an extended 
 infrared source, and therefore can not be used to locate the Brackett line emission.  

\begin{figure}
\plotone{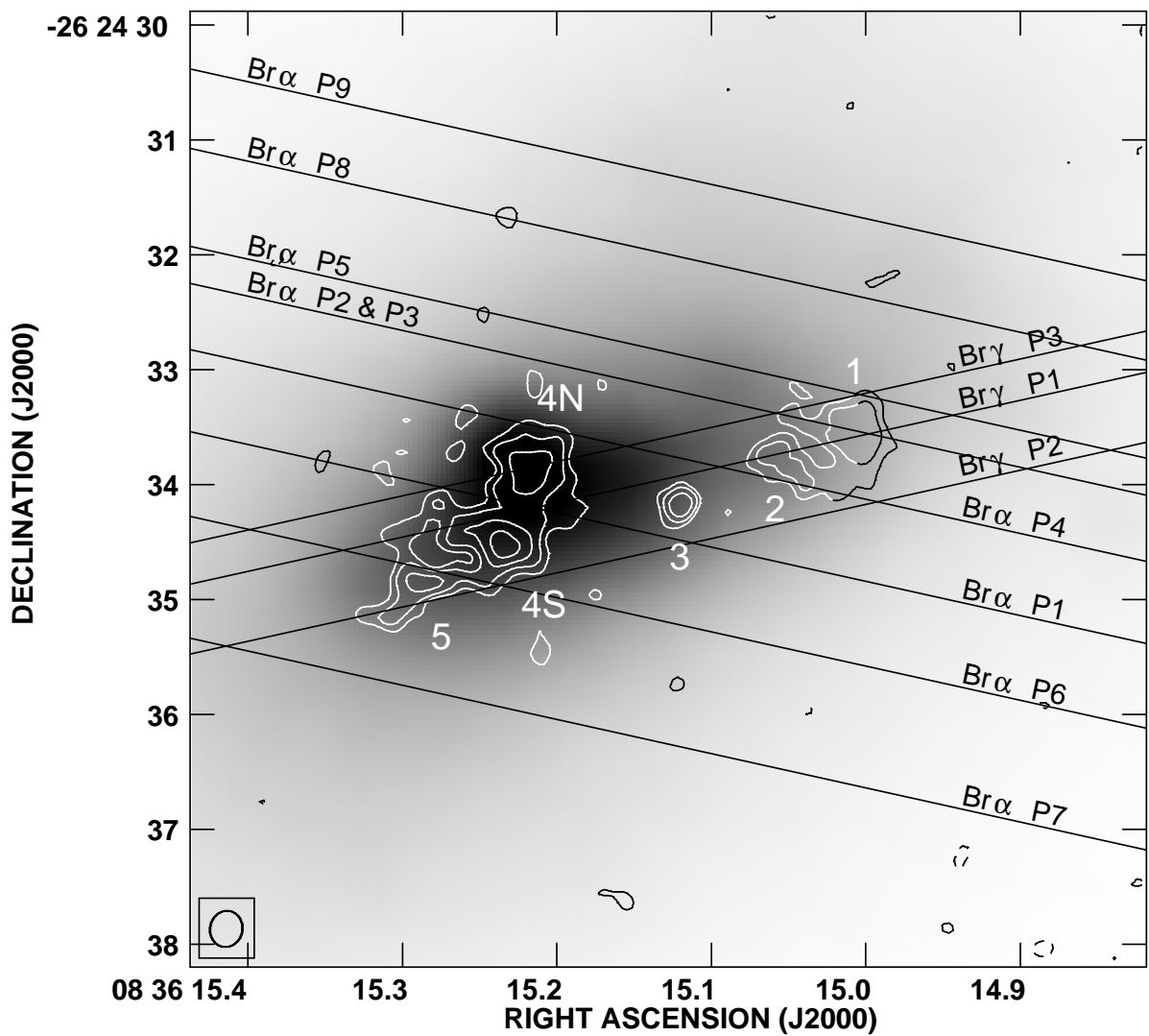}
\caption{A VLA 1.3 cm map (contours), overlaid on an infrared SCAM image (grey scale) 
of He 2-10, with locations of the Brackett line slits.  The coordinates of the SCAM are accurate to $\sim 0.3\arcsec$, 
while the VLA coordinates are accurate to $ \la 0.05$\arcsec.    The 1.3 cm contours are  3$\sigma \times \pm 2^{n/2}$, with
 n = 0, 1, 2, 3 ($\sigma =40~ \mu $Jy beam$^{-1}$ ). The beam has a FWHM of $0.32 \times 0.28"$ , and a p.a. of -12\degr.   Lines mark slit 
 positions of the NIRSPEC Br$\gamma$ and Br$\alpha$ observations.  
  The slits are 0\arcsec.6 wide in Br$\alpha$ and 0\arcsec.4 wide in Br$\gamma$.  For most of the 
  spectral observations, the seeing was larger than the width of the slit.  We detect no spectral line emission 
  from the positions Br$\alpha$ P7 and P9.  K-band continuum is detected in the  Br$\gamma$ spectra of the
   eastern \HII\ complex.  
}
\label{multiplot}
\end{figure}

In Figure \ref{multiplot} we show the central region of an infrared image made 
by coadding the SCAM images from the night of  February 10, 2003, with the slit 
positions superimposed.  He 2-10 was imaged on the SCAM detector in both the 
on-target and off-target spectra.  The images were sky-subtracted, and the dark
 pixels at the location of the slit were added to the bad pixel mask.  No flat field 
 was taken, and no photometric calibration was done.  The SCAM detector is 
 effective between 0.95 and 2.5~\micron, and therefore this infrared image, which 
 was behind the KL filter, has a bandpass of approximately 2.16 - 2.5 \micron. On 
 both nights, the seeing was $\sim 1\arcsec$ and variable, as measured from the 
 point spread function (PSF) of the few stars in SCAM images.         

We were able to determine accurate coordinates for the SCAM image in
 two steps.  First, we created an infrared image which covered enough 
 sky to include a few stars from the Two Micron All Sky Survey (2MASS) 
 by mosaicking SCAM images from both on-target and off-target spectra.
   The resulting coordinates were accurate to $\sim 1\arcsec$,  which is a 
   rather large angular scale compared to the structure in the 1.3 cm map. 
      We then rotated the SCAM image so that north is up and east is to the left, and 
       shifted our coordinate solution by less than 1\arcsec, to align the peak of our
        SCAM emission with the coordinates of the K-band peak from \cite{Cabanac}, which
         is accurate to $\pm 0.3 \arcsec$.    Provided the SCAM coordinates, the slit positions are 
         easily registered with the radio emission (see Figure \ref{multiplot}).

\begin{figure}
\epsscale{0.8}
\plotone {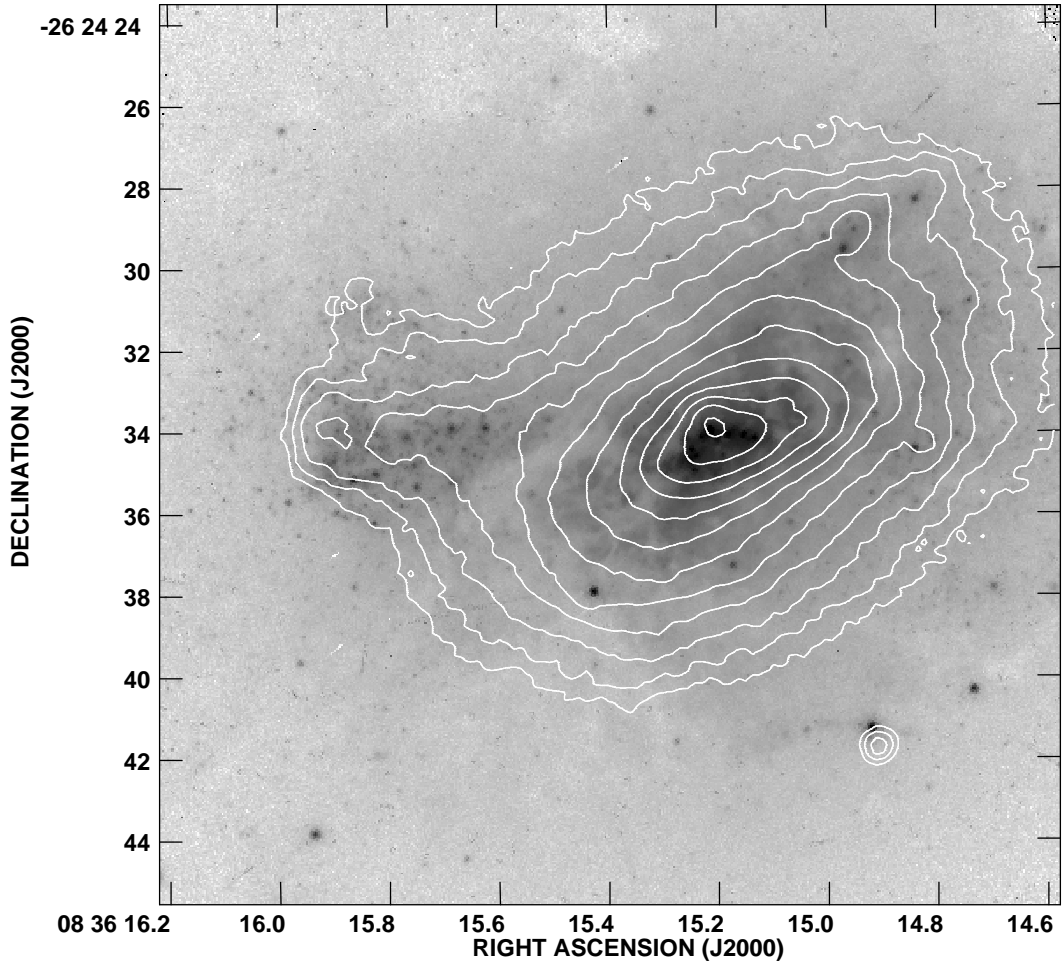}
\caption{An F555W image of He 2-10 (grey scale), with SCAM broad band 2 \micron\ contours.  
The contours are $10\sigma \times 2^{n/2}$.  The extended emission to the east is the optical region B from \cite{Corbin}, and the 
brighter region to the west is region A.  The radio emission in Figure \ref{multiplot} is within the western region.   }
\label{optical}
\end{figure}

In Figure \ref{optical} we show contours of the SCAM emission, in comparison to an F555W image
 taken with the Wide Field Planetary Camera (WFPC2) on the {\it Hubble Space Telescope} (HST). 
  Since coordinates taken from HST are accurate to only 1\arcsec, we take the astrometric solution
   from \cite{Cabanac}, which aligns the peak of the near infrared image with the brightest optical 
   source.  The extended emission to the east is optical region B \citep{Corbin}, while the brighter
    region to the west is region A, where the more active star formation--as reflected by infrared, radio, 
    and nebular line emission-- is underway.

The high resolution spectra were spatially rectified, and a wavelength solution was applied using the 
NIRSPEC reduction package, REDSPEC.   Arc lamp wavelength solutions obtained following these 
procedures are generally accurate to $\sim1 ~{\rm km~ s}^{-1}$ \citep{Prato}.  In the raw L band data, we 
see strong, regularly spaced emission lines, which we attribute to ice on the window of NIRSPEC.  Fortunately, 
these lines are removed by subtracting an image taken in an adjacent piece of sky.  Wavelength calibration was
 done using lamps.   Conditions were unphotometric, and we were unable to estimate slit losses 
 due to variable seeing.

\subsection{VLA Observations \& Calibration}

Archival VLA data, taken in the AB configuration (program AK527), were used to make the 1.3 cm map in 
Figure \ref{multiplot}.   The data were reduced and calibrated using the Astronomical Image Processing 
System (AIPS).   The absolute flux calibrators were 3C48 and 3C286, and the phase
 calibrator was 0836-202.  The measured flux of the phase calibrator was 2.3 Jy, which is within 
 the uncertainties of \cite{JK03} in their original analysis of the data.    The rms in this map is 
 40 $\mu$Jy beam$^{-1}$, and the FWHM of the beam is $0.32 \times 0.28\arcsec$, with a 
 position angle of -12\degr.  We achieve a higher resolution than \cite{JK03} since their 
 objective was to match the beam to longer wavelength observations.  

The clean 1.3 cm map is shown in Figure \ref{multiplot}.  We see the sources organized
 in eastern and western ``clumps'', as well as a central point source. There are six knots
  detected at 5 to 10$\sigma$, and they are labeled, following the identifications in \cite{JK03}.  
    Source 3 is the central point source, and source 4 we resolve into two sources, 
    a northern and a southern source,  which we call sources 4N and 4S.        

In Table \ref{radioflux} we present the flux density of each knot, which we measured using 
AIPS.  The task TVSTAT integrates the flux inside any specified region, and is ideal to
 measure the flux densities of crowded, extended sources, where confusion is an issue.  
 Uncertainties are estimated from multiple attempts to measure the flux  density of each
  knot with TVSTAT, and are $\sim 10-20$\%.   This uncertainty arises mostly from 
  extended emission, which can be difficult to account for.   At 1.3 cm, absolute fluxes
   measured with the VLA  are accurate to $\sim 5$\%.       The flux densities which we
    list in Table \ref{radioflux} are larger than those found by \cite{JK03} because we 
    include more extended emission.

\section{Results}
\label{results}
\subsection{Brackett Line Emission}
\label{brackett_results}

\begin{figure}
\epsscale{1.0}
\plotone{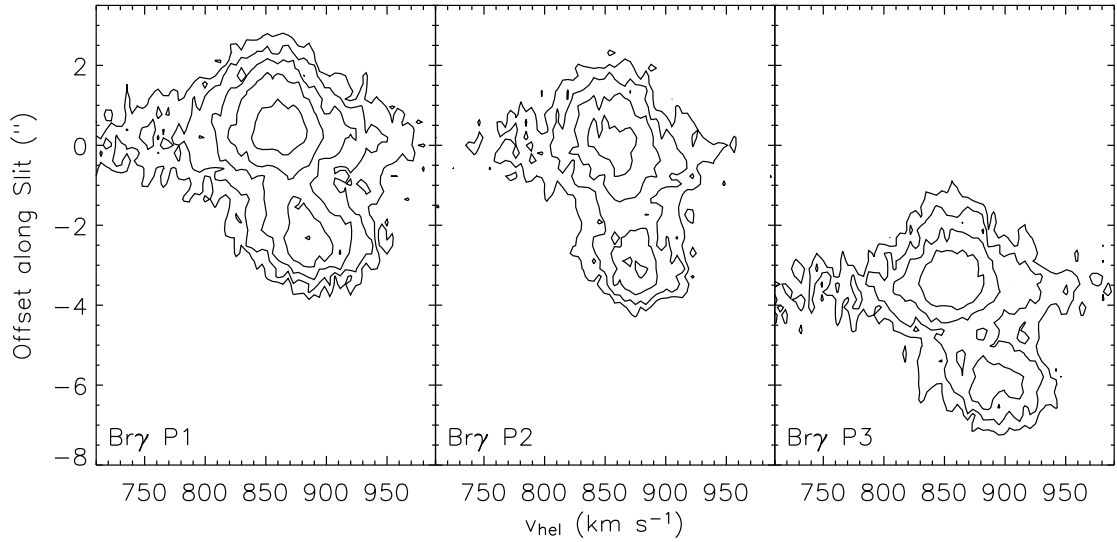}
\caption{Br$\gamma$ two dimensional spectra.  Up is to the southeast, with the top (bottom) peak of 
emission representing the eastern (western) nebulae.  Contour 
levels are 3$\sigma \times \pm 2^{n/2}$, n =0, 1, 2, 3, 4, with $\sigma \approx .06$ counts pixel$^{-1}$.  
   The eastern spectra also display contiuum emission. }
\label{brgcntr}
\end{figure}

\begin{figure}
\plotone{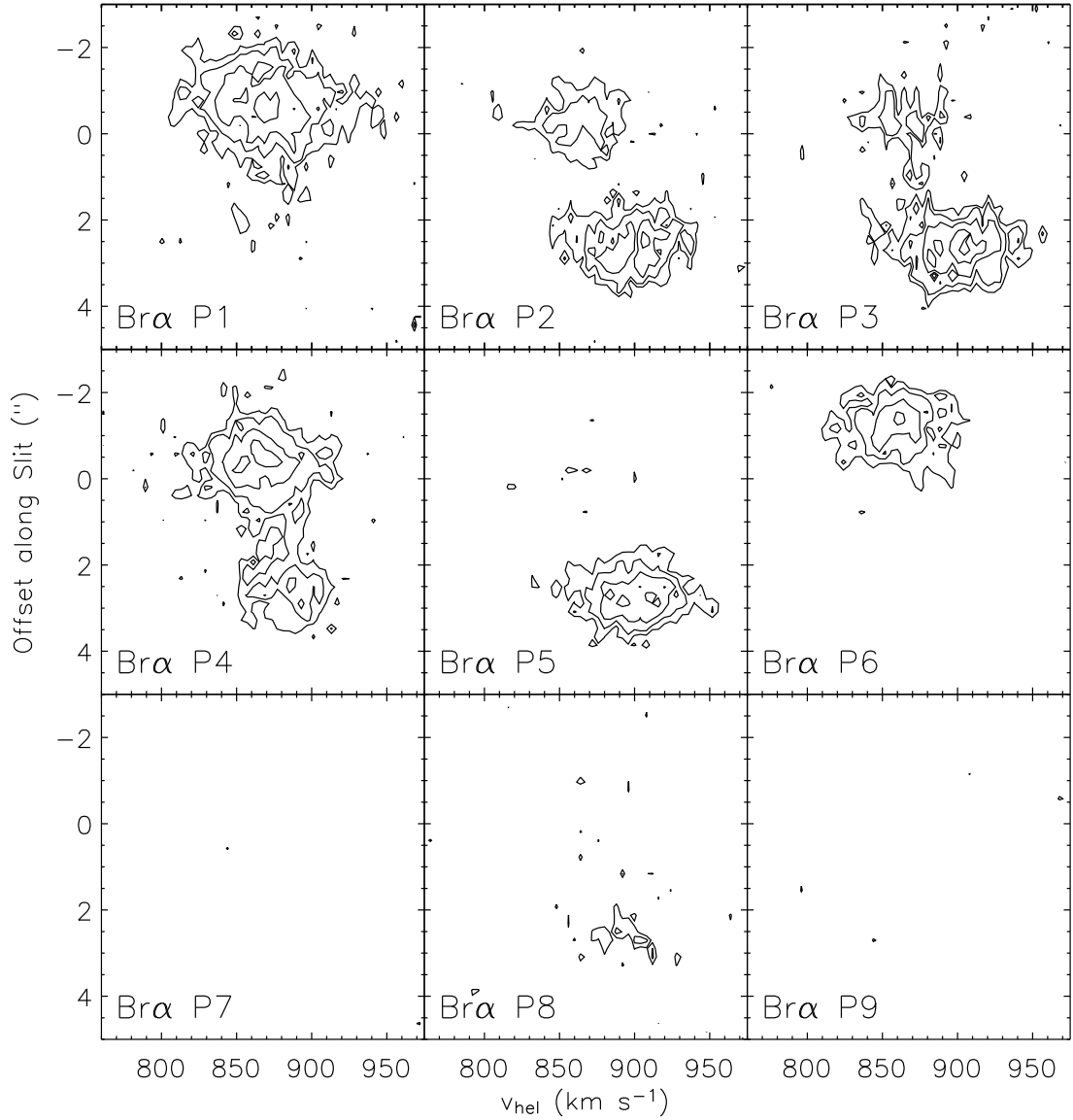}
\caption{Br$\alpha$ two dimensional spectra.  Up is to the northeast, 
with the top (bottom) peak of emission representing the eastern (western) nebulae.  Contour 
levels are 3$\sigma \times \pm~ 2^{n/2}$, n = 0, 1, 2, 3, 4, with $\sigma \approx 0.6$ counts pixel$^{-1}$.
  No continuum sources are detected at any location.    }
\label{bracntr}
\end{figure}

In Figures \ref{brgcntr} and  \ref{bracntr} our two dimensional Br$\gamma$ and Br$\alpha$ spectra 
show  two ``clumps'' of  emission,  separated by approximately 3 - 4\arcsec~along the slit  with an 
obvious velocity offset.  
Although we do not to expect to detect Brackett line emission from radio continuum source 3, 
which is non-thermal \citep{JK03}, we cannot provide an upper limit  because the seeing prevented 
us from resolving source 3 from the eastern and western complexes.  
 K-band continuum emission is detected in the Br$\gamma$ eastern sources, but not the western sources.
  It is strongest in P1, but is apparent in all three spectra.  This is consistent with the 
  SCAM images (see Fig \ref{multiplot}).     There is no continuum emission detected  at 4 \micron.         
   
\begin{figure}
\plotone{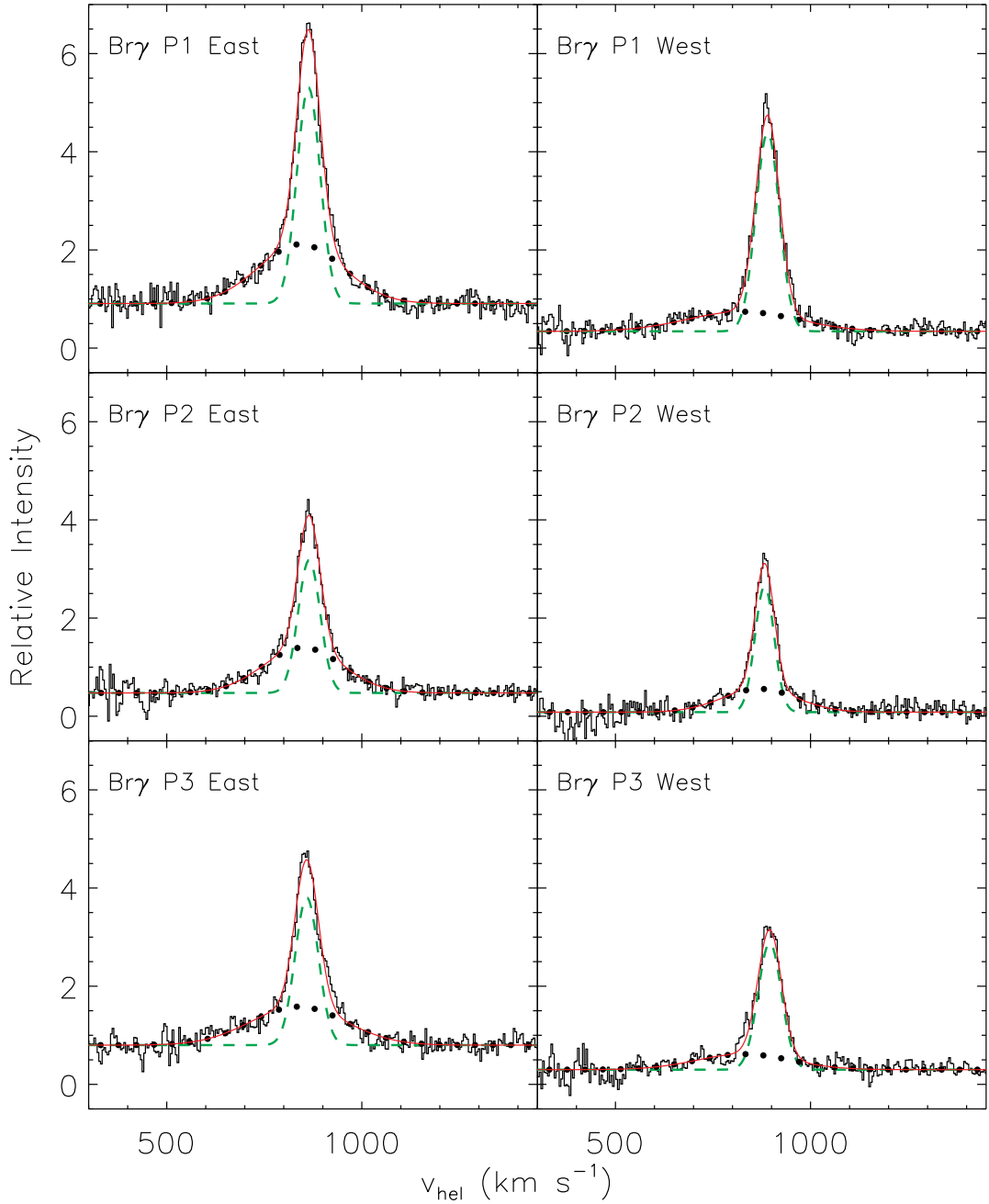}
\caption{Br$\gamma$ 1-D Spectra. These spectra are made by integrating
 over 1.4\arcsec~ along the slit, centered on the peaks of emission which are 
 present in Figure \ref{brgcntr}.  The red curve shows our two component
  Gaussian fit to the spectral lines.  The dashed green curve is the narrow
   component of the fit, and the dotted curve represents the broad component.    }
\label{brgspec}
\end{figure}

In Figures \ref{brgspec} and \ref{braspec} we show the 
Br$\gamma$ and Br$\alpha$ spectra, obtained by summing over a 
 1.4\arcsec\ area along the slit, centered on the peaks of the eastern and 
 western regions.   Since the seeing was 1-1.4" over the observations, little or
  no spatial information was lost.   In the Br$\gamma$ lines, the signal to
   noise is high enough to distinguish a broad, low intensity blue line wing. 
    To quantify this component, we fit all of the Br$\gamma$ spectral lines 
    with the superposition of two Gaussian curves, as shown in Figure \ref{brgspec}. 
     Line centroids and FWHM velocities of the line profiles, as well as the characterizations 
     of the broad and narrow components,  are listed in Table \ref{kinematics}.  
     The average velocity of the eastern sources is 862 km s$^{-1}$,  while the average 
     western velocity is 889 km s$^{-1}$, giving an average offset  of 27 km s$^{-1}$ between
      the two regions.  For comparison, \cite{Mohan} find $v_{hel} = 873 ~{\rm km~ s}^{-1}$ from their 
      H92$\alpha$ radio recombination line observations, which, at a 
      resolution of $190\times220$ pc, average over the structure that we resolve spatially with NIRSPEC.              
  
 In addition to line profile observations, we consider relative photometry of 
 eastern and western sources, using the  Br$\gamma$ P1 spectrum where
  the slit went approximately through the center of each region.  We find that, 
  of the total flux in this echellogram, $\sim$ 35\% is in the western sources, and 
$\sim$ 65\% is in the eastern sources.  This, agrees well with the relative 
fluxes in the radio, as well as at 11.7 \micron~  \citep{Beck01}.  On the other hand, 
Figure \ref{multiplot} shows less near infrared continuum emission coming from
 the western sources.  This is noted by \cite{Cabanac}, who find that extinction is larger
  near the western complex of \HII\ regions.  

\subsection{Radio continuum}
\label{radio_results}
In Table \ref{radioflux} we present our measurements from the 1.3 cm map, which
 we displayed in Figure \ref{multiplot}.  We measure 4.05 $\pm$ 0.36 mJy from the 
 eastern sources, 1.94$\pm$ 0.15 mJy from the western sources, and 0.40 $\pm$ 0.08 mJy  
 from the central source. The flux densities measured here are consistent with  \cite{JK03}, in
  their original analysis of the data. No single dish observations are available for comparison at
   1.3 cm, but \cite{KJ99} report a 2 cm flux density of  $21 \pm 1.2$ mJy and a 2 cm to 6 cm spectra
   l index of $-0.62\pm 0.11$ from low resolution VLA data.  From this, we infer a total 1.3 cm flux density
    of $16.6 \pm 1.2$ mJy, which indicates that some extended 1.3 cm emission is missed 
    in the data presented here.

 Much can be inferred from radio continuum measurements.  First, one can determine the ionization
  rate from the thermal free-free emission: 
\begin{equation}
\label{nlyc}
\left( {N_{Lyc} \over s^{-1} } \right) \ga 9 \times 10^{49}  \left( {T_e} \over {8000~ K } \right)^{0.35}  \left( { \nu \over {\rm 22~ GHz}} \right)^{0.1}
 \left( {D \over {\rm Mpc}}\right)^2 \left({F_{\nu} \over {\rm mJy}}\right).  
\end{equation}    
   Since He 2-10 has solar metallicity
    \citep{Johansson}, we use $T_e = 8000$ K, typical of Galactic \HII\ regions.    
  With the 1.3 cm flux densities which we measured, we 
  derive $N_{Lyc} =   3.0 \times 10^{52}$ s$^{-1}$,  and $N_{Lyc} =   1.4 \times 10^{52}$ s$^{-1}$, for
   the eastern and western regions, respectively.    
  These estimates of the ionization rate are lower limits, since they 
  assume that the \HII\ regions are optically thin and ionization bounded.  
Flux densities and brightness temperatures of the individual clusters are 
shown in Table \ref{radioflux}.  

With the exception of source 3, the clumps in the 1.3 cm map, which we show in 
Figure \ref{multiplot}, appear to be extended.  Assuming a single component Gaussian source,
 we use the AIPS task IMFIT to find the FWHM of each cluster.  These sizes  range
  from  30-40 pc for sources 1, 2, 4N and 4S. Source 5, which is significantly extended
   in the east-west direction, displays a major axis of 80 pc, and a minor axis of 30 pc.  
   For comparison, the FWHM of the beam is 13 pc.  These sources are obviously extended. 

We also use the brightness temperature of the knots to estimate the nebular 
densities and some spatial information.  For an optically thick \HII\ region which 
fills the beam, we expect $T_b = 8000$ K.  \cite{JK03} find that the flat spectra 
of the thermal sources turn over and become optically thick between 6-14 GHz,
 indicating that  at 1.3 cm, the sources are optically thin. Here, the range of values 
 represent the uncertainty in the turnover frequency, and not  properties of the
  individual knots, which are similar.  Correcting for optical depth, we 
  expect $T_b \sim 350-2000$ K for sources that fill the beam for this range 
  of turnover frequency.    If we compare the observed $T_b$ of Table \ref{radioflux} to 
  this predicted  $T_b$, we obtain a beam filling fraction of 0.005 - 0.03, corresponding to 
  sources of 0.4-1 pc in size.  We infer that each radio source must contain further unresolved 
  clumps of nebular material that do not fill the beam.

 The turnover frequency of 6-14 GHz implies an emission measure, 
 $\int n_e^2 dl  \sim 1-6 \times 10^8$ cm$^{-6}$ pc.   The density is not well 
 constrained, as the turnover frequency and source sizes are both uncertain.   
 The radio nebulae excited by clusters of young stars are usually found, when 
 observed with the highest VLA resolution, to have sizes 1- 10 pc (\citealt{TB04}; \citealt{Tsai}),
  which is consistent with the intensities seen here.   If we adopt these as 
  typical sizes $n_e$  ranges from $10^3$ to a few $\times 10^4$ cm$^{-3}$.   
Estimates of nebular mass are uncertain because the density and geometry are not known.  
 If the sources are of order 1 pc, and the density $10^4$ cm$^{-3}$, the mass of the 
 ionized gas in one spherically symmetric nebula is $\sim 1000~ M_{\sun}$.  Lower densities 
 are possible, and multiple nebulae are more than likely, so the nebular mass 
 associated with each radio source can be expected to lie between a few hundred and 
 several thousand solar masses.     

\begin{figure}
\epsscale{0.8}
\plotone{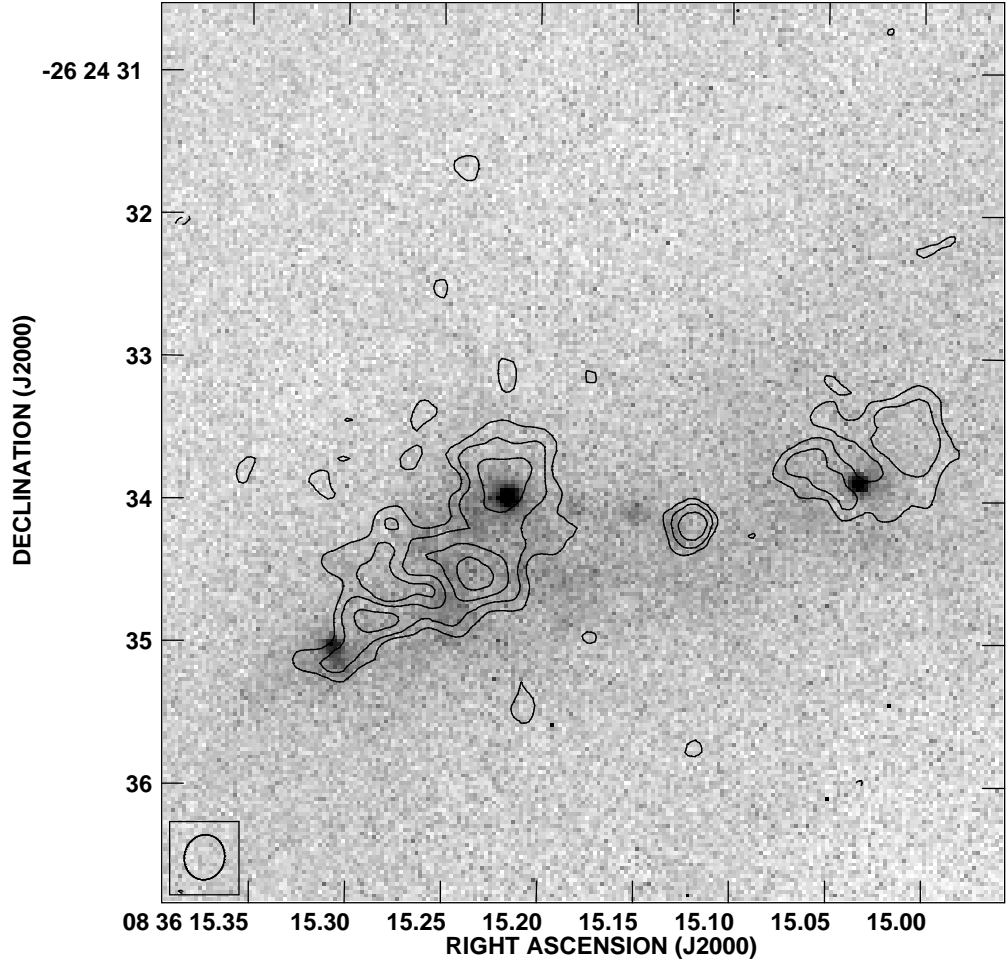}
\caption{ {\it Grey scale:} L\arcmin\ image (courtesy of \citealt{Cabanac}).  {\it Contours:} VLA 1.3 cm map. 
 The contour levels are 3$\sigma ~\times~ \pm 2^{n/2}$, with n = 0, 1, 2, 3.  The radio beam, 
 shown in the lower left, is $0.32 \times 0.28 \arcsec$, with p.a.  = -12\degr.   
 The positional uncertainty associated with the L' image is 0\arcsec.3, so it is uncertain which 
 infrared sources, if any, correspond to the radio knots.  In fact, there is no possible shift which 
 will perfectly align the three brightest infrared sources with radio knots.  }
\label{Lprime}
\end{figure}

 In Figure  \ref{Lprime} we show an L\arcmin\ map, which we received, courtesy 
 of \cite{Cabanac}, with our 1.3 cm contours overlaid.  The positional uncertainty
  associated with the L\arcmin\ image is 0\arcsec.3, which is still large compared to
   the expected size of super star clusters ($\sim$ 1 pc).  Since each radio knot is likely 
   a complex of smaller nebulae, the infrared sources could correspond to any unresolved 
   clump or multiple thereof.    As a check, we tried a variety of alignments between the 1.3 cm
    and the L\arcmin\  image, and under no circumstances were we able to simultaneously align
     the three brightest infrared sources (L1, L4a, L5 from \citealt{Cabanac})  with radio knots. 

\section{Discussion}
\label{discussion}
\subsection{UV Continuum Photons}
\label{uvphot}
In \S \ref{radio_results} we calculated a total Lyman continuum rate 
of N$_{Lyc} = 5.3 \times 10^{52}$ s$^{-1}$, from the radio continuum map, 
 for $T_e = 8000$ K.   A number of sources have reported the  ionization rate 
 from different infrared and radio diagnostics 
 (\citealt{Mohan}; \citealt{Beck01}; \citealt{Johnson00}; \citealt{KJ99}; \citealt{JK03}),
  and a range of values have been derived.   Before we attempt to determine the O 
  star content in He 2-10, we must first account for the different estimates of N$_{Lyc}$.  Here,
   we compare measurements which have been made in the infrared or radio, where the 
   effects of extinction are negligible, and the light from the youngest, embedded 
   sources can escape.

We begin by considering the Brackett line fluxes, which were
 reported by \cite{Kawara89}, F$_{Br\alpha} = 2.6 \times 10^{-16}$ W m$^{-2}$, and
  F$_{Br\gamma} = 4.4 \times 10^{-17}$ W m$^{-2}$.   \cite{HBT} summarized the
   usage of Brackett lines as in indicator of star formation, and provide tables of the
    intrinsic Brackett line intensity ratio, as well as the ratio of N$_{Lyc}$/N$_{Br\alpha}$, 
    based on the tables of \cite{brocklehurst}, \cite{Giles}, and \cite{HS}.   
    From the fluxes of \cite{Kawara89},  we obtain N$_{Lyc} = 5.7 \times 10^{52}$ s$^{-1}$ assuming 
    no extinction is present at Br$\alpha$.
However, we can correct for extinction at Br$\alpha$ using 
the differential extinction between Br$\alpha$ and Br$\gamma$.  
We use a $\lambda^{-\beta}$ emission law, with $\beta =  1.5$, which 
gives A$_{\alpha} = 0.48$, and $A_{\gamma} = 1.2$.  With extinction correction, 
we get  N$_{Lyc} = 6.9 \times 10^{52}$ s$^{-1}$.  

Other radio continuum measurements have been made by  \cite{Beck01}, \cite{KJ99}, and \cite{JK03}.  
From 2 cm VLA data, \cite{KJ99} measured a total N$_{Lyc} = 3.5 \times 10^{52}$ s$^{-1}$ by fitting 
Gaussian sources to the compact cores.  From the same data,  \cite{Beck01}  found a total  
$N_{Lyc} =  7.3 \times 10^{52}$ s$^{-1}$, by including the extended emission as well as the compact sources.  

In Table \ref{nlyct} we summarize these results.  We 
find that the 2 cm measurement of \cite{Beck01} agrees
 with our estimate from the extinction corrected Br$\alpha$ luminosity.  This 
 suggests that the 1.3 cm measurements presented in this paper, and the 7 mm 
 measurements of \cite{JK03} have resolved out some of the thermal emission.  Considering 
 the previous measurements,  we estimate   $N_{Lyc} \ga 7 \times 10^{52} s^{-1}$.   This means that 
 the total ionizing stellar content is equivalent to $\ga$ 7000 O7 stars, where  one O7 star produces 
 10$^{49}$ photons s$^{-1}$.        
   
 \subsubsection{Cluster Masses \& Luminosities }
 \label{masses}
  We next use the population synthesis models of \cite{Leitherer} to estimate the stellar mass 
   and luminosity from the ionization rate.  
  While the age of the starburst is generally important, the Lyman continuum rate is
   insensitive to age until the O stars evolved off of the main sequence and the first supernovae
    occur.  For an upper mass cutoff of $35~M_{\sun}$ \citep{Beck97}, this does not happen until an
     age of about 7 - 8  Myrs \citep{Leitherer}.  The clusters appear to be younger than this, 
     so for estimating mass, we assume that they are at the Zero Age Main Sequence (ZAMS) stage.      
 We also assume a solar metallicity, and a 
 Salpeter initial mass function (IMF) ($\alpha  = -2.35$), with a lower mass cutoff of 
 1 $M_{\sun}$.  We find that a luminosity of $3\times 10^9 L_{\sun}$, and total stellar mass of at 
 least $6 \times 10^6~M_{\sun}$ will explain the inferred Lyman continuum rate for the entire starburst.      If the IMF 
 extends down to $0.1~M_{\sun}$, then the stellar mass will be more than twice as large.  This  mass 
 agrees well with the value derived by \cite{Beck97}, using different models, but also using the Brackett 
 line fluxes of \cite{Kawara89}.  On the other hand, our derived mass is approximately six times larger 
 than the mass determined in \cite{JK03}, due to their choice of an upper mass cutoff of 100~$M_{\sun}$.
   This is an indication of the uncertainty in mass estimated from $N_{Lyc}$ using different IMFs.   In
    Table \ref{radioflux}, we list the individual source masses, which range
     from $\sim 1 - 2 \times 10^{6}~ M_{\sun}$.     


\subsection{Kinematics}
\label{kin}
\subsubsection{The Dynamical Center: Star Formation in a disk? }
How do the velocities we observed with NIRSPEC, which measure the motions 
of the \HII\ regions of the starburst, relate to the kinematics of the galaxy? The kinematics of 
He 2-10 are complex: it  is an advanced merger, with a tidal tail extending to the 
southeast (\citealt{K95}; \citealt{Meier}), and extensive dust lanes which can frustrate 
attempts at optical identification of structure.  \cite{K95} mapped the galaxy in CO and \HI\, and they 
find that the CO may be filling a hole in the \HI\ distribution, much like large spiral galaxies with nuclear condensations 
of molecular gas (\citealt{Scoville}; \citealt{ML78}).  Since the \HII\ regions are roughly coincident with the 
CO emission, they are likely within the \HI\ hole.

In addition, the \HI\ observations show rotation with a systemic velocity
 of $873 \pm 20~ {\rm km ~s}^{-1}$ (heliocentric).  The central CO emission also displays 
  a sense of rotation which is consistent with the larger scale \HI\ emission, even though the  
  CO has an asymmetric, tidal tail appearance.   We find velocity centroids of $890 \pm 10 ~{\rm km~ s}^{-1}$ for the 
  western \HII\ regions, and $862 \pm 5 ~{\rm km~ s}^{-1}$ for the eastern \HII\ regions.  We compare 
  these velocities to the CO channel maps in \cite{K95}, and find that the western \HII\ regions are 
  approximately coincident with the peak of the CO in their $896~{\rm km~s}^{-1}$ channel, and the 
  eastern \HII\ regions are close to the peak of the CO emission in their $865 ~{\rm km~s}^{-1}$ channel.  This is 
  consistent with the \HII\ regions moving at the velocities of the molecular clouds from which they were born.

\subsubsection{Line Profiles: Width} 
We present the measured line widths of all the lines in Table 1. The FWHM are 
consistently 60-80 km s$^{-1}$.      The widths of hydrogen recombination lines from a \HII\ regions will 
include the effects of thermal broadening, electron impact broadening, and broadening due to different 
velocities in the gas (added in quadrature).   Thermal broadening at $T_e$ of 8000 K is $\sim$ 20 km s$^{-1}$ (FWHM),  and 
electron impacts have a negligible effect at the low quantum numbers of Brackett lines.   So the non-thermal width of the lines, 
that is the width due to gas motions, is $\sim$57 to 77 km s$^{-1}$.

Are the recombination lines in He 2-10 wide because at 9 Mpc, the beam encompasses more 
sources than in Galactic surveys?  We can estimate the contribution to the gas motions of the 
velocity dispersion of the multiple sources in the beam.  Based on the estimated masses of the 
star clusters within the nebulae (\S \ref{uvphot}) of 4 and $2\times 10^{6}~ M_{\sun}$, and for the 
observed sizes of 50-70 pc,  we would predict Brackett linewidths of $15-20 ~{\rm km ~s}^{-1}$ for both 
the eastern and western complexes.  
Since the contributions of the gas motions and the thermal line width add in quadrature,  there must 
remain a  component of 51-75 km s$^{-1}$  in the line widths which must be due to turbulence, outflow, 
or some other velocity field in the gas.

The observed line widths are highly supersonic, but not at all unusual for dense, young
star formation regions.  \cite{GL99} tabulate the widths 
of recombination lines from Galactic compact \HII\ regions and find that 
while most have FWHM between 30 and 50 km s$^{-1}$,  many have 
FWHM as high as 80 km s$^{-1}$.  However, these Galactic \HII\ regions 
are excited by single stars and do not have the velocity dispersion contribution
 of the clusters in He 2-10. 
 These core line widths for \HII\ regions with luminosity $\sim 10^9 L_{\sun}$, 
 are similar to those of Galactic regions with luminosity $\sim 10^5 L_{\sun}$ \citep{Turner03}.  

 If each of our radio knots is treated as a single nebula, 
 expanding at 30-40 km s$^{-1}$ to its current size, they would be $3-7\times 10^5$ years old.  
 But  Galactic Compact \HII\ regions under the same analysis would be only a few thousand
  years old: those dynamical arguments greatly underestimate the age of \HII\ regions (\citealt{DW81}; \citealt{WC89}).  

Because the individual clusters within each radio knot are unresolved
 we cannot calculate the escape velocities.  But given the large stellar masses,
  and the small radii observed for other sources, such as the one in the 
  dwarf galaxy NGC 5253 \citep{Turner03} it is possible that for at least some
   of the sub-sources the escape 
velocities are comparable to the supersonic gas velocities observed. If this is so, 
then gravity will play an important role in the evolution of these nebulae and, unlike the case 
in typical Galactic \HII\ regions, cannot be neglected.

\subsubsection{Line Profiles: High Velocity, Blue Wings}
\label{wings}
\begin{figure}
\epsscale{0.8}
\plotone{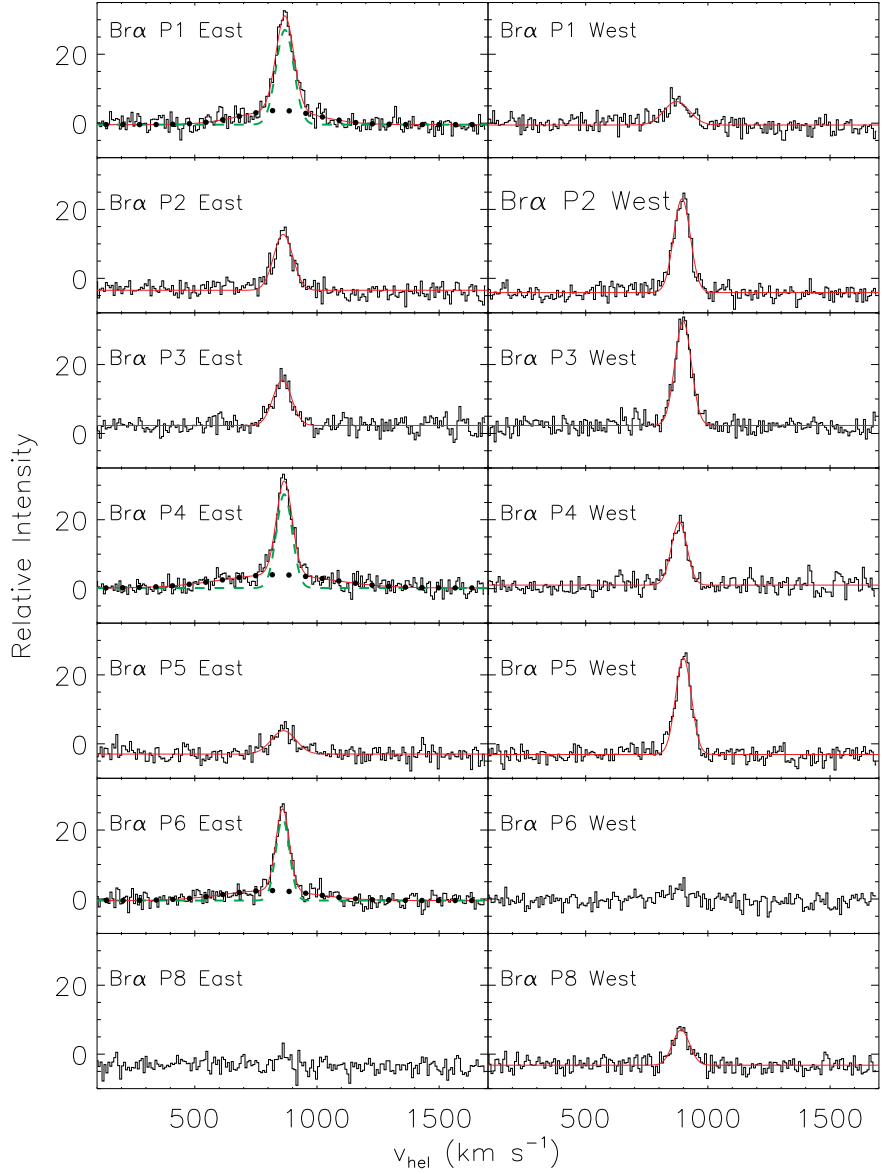}
\caption{Br$\alpha$ one dimensional spectra.  These spectra are made
 by integrating over 1.4\arcsec~ along the slit, centered on the peaks of emission 
 which are present in Figure \ref{bracntr}.  The red curve shows our Gaussian fit to the 
 spectral lines.  In some cases we fit two component Gaussians, which are shown as green
  dashed curves (narrow) and dotted curves (broad).  For these spectral lines, the red curve is the
   sum of the two components.   }
\label{braspec}
\end{figure}

In Figure \ref{brgspec}, one of the most obvious spectral line features
 is the asymmetric, blue line wing, which is most obvious in the Br$\gamma$ P1 spectrum.  
 This feature is also seen in H$\alpha$ line profiles (\citealt{Kawara87}; \citealt{Mendez97}).   A single 
 Gaussian profile is inadequate at describing the line profile, so we fit the line with the superposition of two 
 Gaussian curves, as described in \S \ref{brackett_results}.  The broad, low intensity component 
 is shifted approximately 20 km s$^{-1}$ blue of the central peak.  For comparison, we fit similar profiles
  to some of the Br$\alpha$ spectra in Figure \ref{braspec}, where a similar blue line wing
   is detected at $\sim 1.5 - 2.0 \sigma$.  We list the parameters of these fits in Table
    \ref{kinematics}.  For Br$\gamma$, the FWHM of the broad component is  
    200-300 km s$^{-1}$, which is consistent with H$\alpha$ profiles presented 
    by \cite{Kawara87}.  

We consider possible explanations for the high velocity gas motions.
First, He 2-10 contains several optical clusters which display 
WR features \citep{Chandar}, and do not have strong Brackett-emitting counterparts.  There
 are also large scale bubbles present in H$\alpha$ narrow band images, and similar 
 high velocity motions are present in H$\alpha$ lines, out to distances of $\sim$ 500 pc 
 from the starburst (\citealt{Mendez97}; \citealt{Mendez99}).  One possibility, is 
 that one or several of the centrally located, more evolved optical clusters is driving 
 high velocity gas motions.  In this model, stellar winds impinge on nearby, 
 Brackett-emitting \HII\ regions, causing the high velocity motion which we see.  
   We note that  the broad component is stronger in the eastern regions than the 
   western regions, as seen in Figures \ref{brgspec}  and \ref{braspec}.
This could be the result of higher extinction in the western regions, or there may 
simply be more high velocity gas in the east.

Alternatively, the embedded, young stars within the Brackett line emitting 
regions could be the source of high velocity motions.  
 \cite{Bouret}  and \cite{Puls} have modeled H$\alpha$ line profiles of 
 O star winds in the Galaxy, and for sources where the P-Cygni absorption is 
 minimal or non- existent, they exhibit similar, asymmetric line profiles.   In these 
 sources internal obscuration suppresses the red wing of the line, causing asymmetry.
   \cite{Kawara89} suggest an average of 1-2 magnitudes of extinction in the K-band, so this is not 
   unreasonable.  In a stellar wind, we expect this effect of extinction to be less 
   pronounced in Br$\alpha$, resulting in a broad component which is wider, and
    redshifted relative to that in Br$\gamma$.  
 Where the broad component is detected, it has a FWHM of 300-500 km s$^{-1}$, significantly 
 wider than the broad Br$\gamma$ components.  However, we can not determine if the 
 centroid of the broad Br$\alpha$ component is shifted  because of the low S/N in the
  wing of the Br$\alpha$ line and the differences in the slit orientation between Br$\alpha$ and Br$\gamma$.  
 
In actuality, we are observing \HII\ regions rather than stellar winds, 
and complex kinematics are a reality.  \HII\ regions have a variety of morphologies, including 
champagne flows, shells, or cometary structure (see \citealt{GL99}).  It is possible 
that higher velocity gas is more deeply embedded in the nebulae, and that  Br$\alpha$ is mapping 
different kinematics than the Br$\gamma$.  This could explain the larger velocities in the broad component of Br$\alpha$, 
without requiring a redshift relative to Br$\gamma$.

\section{Conclusions}
The dwarf galaxy He 2-10 contains an intense starburst, which 
is resolved into two major complexes of \HII\ regions, separated by  approximately 150 pc.  
What has triggered the starburst, and why are there two complexes of \HII\ regions?  Is there anything 
we can infer about the evolution of the starburst?  The high extinction and dense gas in the \HII\ regions 
indicate that these star-forming regions are young.  The \HII\ regions are both spatially and kinematically 
coincident with the CO  emission (\S \ref{kin}).    Whether the fundamental cause of the starburst is a 
merger or accretion of interstellar gas, the immediate cause is a concentration of molecular gas near 
the center of the galaxy.

We have analyzed high spectral resolution Brackett line spectra of 
the two \HII\ complexes in Henize 2-10. The Brackett line profiles provide substantial 
kinematic information about the star clusters and associated nebulae.  The line 
profiles, particularly Br$\gamma$ where the S/N is high, show a broad blue wing.  The  line 
profile can be fit by a two component Gaussian, where one component is broad, low intensity,
 and blue shifted relative to a more narrowly peaked line center.  This line profile can be 
 explained by extinction internal to the nebulae, which is suppressing the emission from 
 gas that would otherwise appear redshifted.  This high velocity gas motion, taken with 
 the clumpy nature of the sources, suggests that the young star clusters in He 2-10 may be in the process 
 of dispersing their nebular material via winds.

The two complexes of \HII\ regions appear to be associated with 
two kinematically distinct molecular clouds.  
 We find an obvious velocity offset between our eastern and western \HII\ regions.  Comparison
  to the CO channel maps of \cite{K95} suggests that the Brackett line centroids are consistent with 
  the velocities of the molecular gas from which these clusters formed.  

We also use archival VLA observations of the \HII\ regions to estimate stellar mass, 
as well as the sizes and nebular densities of the \HII\ regions.  We find that each thermal 
radio source contains  $1-2 \times 10^{6}~M_{\sun}$.  The 1.3 cm maps also indicate 
that the mass and nebular material are not distributed in a single cluster and nebula.  There are 
most likely multiple clumps of star formation associated with each radio knot.  That the \HII\ regions appear 
in the radio continuum and are not extremely optically thick suggests that they are somewhat evolved and
 beginning to expand and disperse.

The line profiles we see in He 2-10 present the same paradoxes seen in NGC 5253 and other galaxies 
(\citealt{Turner01}; \citealt{Turner03}).  The He 2-10 clusters contain many O stars, each of which may be 
presumed to drive a wind with velocity on the order of 1000 km s$^{-1}$.  Yet the narrow core of  the He 2-10
 lines (FWHM) is no wider than that  of many Galactic  \HII\ regions with one O star. 
Unlike the NGC 5253 source, the He 2-10 lines have weak but broad wings.  
This shows a body of fast moving gas that may soon break out of the original cluster environment, 
while much of the line luminosity is confined.  

The picture of He 2-10 presented by these kinematical considerations is a snapshot of an aging starburst, 
in which the neutral gas is beginning to settle into patterns typical of spiral galaxies, with centrally concentrated 
molecular gas, and an \HI\ ring.  The starburst is aging, and the winds of the young stars are becoming prominent in 
the ionized gas, reflected in blue-shifted flows of FWHM 200-300 km s$^{-1}$.  Discovering the origins of the starburst 
amid these intense evolutionary forces will be challenging.

\acknowledgments
The authors wish to recognize and acknowledge the 
significant cultural role and reverence that the summit of Mauna Kea 
has always had within the indigenous Hawaiian community.  We are fortunate to
 have the opportunity to conduct observations from this mountain.  
We wish to thank R. Cabanac for providing L\arcmin\ data.  This work
 was supported in part by NSF grant AST 0307950.

{\it Facilities:} \facility{Keck:II (NIRSPEC)}, \facility{VLA}

\clearpage
               
\clearpage

\clearpage

\clearpage

\clearpage 

\clearpage

\clearpage

\clearpage

\begin{deluxetable}{ccccccc}
\tabletypesize{\footnotesize}
\tablecaption{Radio Knot Properties}
\tablewidth{0pc}
\tablecolumns{7}
\tablehead{
\colhead{Source} & \colhead{$\alpha$}  & \colhead{$\delta$} & \colhead{$F_{1.3 cm} $} & \colhead{$T_B$ \tablenotemark{b}}&   \colhead{O7 stars\tablenotemark{c}} & \colhead{Mass\tablenotemark{c}  }   \\ 
\colhead{} 	     & \colhead{(J2000)} &  \colhead{(J2000)}& \colhead{(mJy)}    & \colhead{(K)} & \colhead{} & \colhead{($10^6 M_{\sun}$)} \\
}
\startdata
1	&  8 36 15.000 & -26 24 33.600  & 1.08 $\pm$ 0.11\tablenotemark{a}   & 7.0   &   1900 & 1.3  \\
2	&  8 36 15.062 &-26 24 33.720  & 0.86 $\pm$ 0.10   & 6.4   &  1100 & 0.7 \\
3 \tablenotemark{d} & 8 36 15.122 & -26 24 34.185 &  0.40 $\pm$ 0.08   & 10.0  &  \nodata  & \nodata   \\
4N & 8 36 15.218 & -26 24 33.905 &  1.61 $\pm$ 0.25   &  9.4   & 2300  & 1.5   \\
4S & 8 36 15.235 & -26 24 34.960 & 1.52 $\pm$ 0.24   &  12   & 2500 &  1.7    \\
5 & 8 36 15.286 &-26 24 34.840 &	  0.92 $\pm$ 0.10   &   8.5   &  1200 & 0.8  \\
\enddata
\tablenotetext{a}{Uncertainties are estimated from multiple attempts to measure the flux density of each knot with TVSTAT. }
\tablenotetext{b}{1.3 cm brightness temperatures measured with the VLA are accurate to $\pm$5 \%}
\tablenotetext{c}{Assumes a Salpeter IMF with $1 ~M_{\sun} \le M \le 35 ~M_{\sun}$, and a total $N_{lyc} = 7\times10^{52} ~{\rm s}^{-1}$, with resolved out emission distributed equally among the sources.}
\tablenotetext{d}{\cite{JK03} show that this source is nonthermal.}
\label{radioflux}
\end{deluxetable}

\clearpage

\begin{deluxetable}{cccccccccccccccc}
\tabletypesize{\scriptsize}
\rotate
\tablecolumns{16}
\tablewidth{0pc}
\tablecaption{Brackett Line Profiles}
\tablehead{
\colhead{} & \multicolumn{7}{c}{Eastern Region} & \colhead{} & \multicolumn{7}{c}{Western Region} \\
\cline{2-8}  \cline{10-16} \\
\colhead{} & \multicolumn{2}{c}{Total} & \multicolumn{2}{c}{Narrow} & \multicolumn{2}{c}{Broad} & \colhead{F$_{narrow}/F_{broad}$ } &\colhead{}   & \multicolumn{2}{c}{Total} & \multicolumn{2}{c}{Narrow} & \multicolumn{2}{c}{Broad} &\colhead{F$_{narrow}/F_{broad}$}   \\
\colhead{Spectrum} & \colhead{$v$} & \colhead{FWHM} & \colhead{$v$} &\colhead{FWHM}   & \colhead{$v$} &\colhead{FWHM} & \colhead{} & \colhead{} & \colhead{$v$} &\colhead{FWHM}& \colhead{$v$ } &\colhead{FWHM}& \colhead{$v$} &\colhead{FWHM}  & \colhead{}  \\
\colhead{} & \multicolumn{6}{c}{(km s$^{-1}$)} & \colhead{} &  \colhead{} & \multicolumn{6}{c}{(km s$^{-1}$)} & \colhead{}  \\ 
\cline{2-7} \cline{10-15} \\
\cline{1-16}
\multicolumn{16}{c}{Br$\gamma$}
}
\startdata
 P1   &  861$\pm 1$  & 66$\pm 2$ & 863$\pm 1$  & 70$\pm 3$ &   843 $\pm 3$ & 254$\pm 10$  & 1.0$\pm 0.1$   &    & 887$\pm 1$  & 61$\pm 2$ &  890$\pm 1$  & 71$\pm 1$ &824$\pm 11$  & 331$\pm 24$&  2.2$\pm 0.2$ \\ 
P2   & 860$\pm 1$   & 62$\pm 2$  & 865$\pm 1$ & 67$\pm 2$ &   849$\pm 4$ & 237$\pm 13$   & 0.8$\pm 0.1$ &    & 880$\pm 1$  & 57$\pm 2$&  882$\pm 1$  & 57$\pm 2$  & 869$\pm 7$ & 223$\pm 21$ &  1.4$\pm 0.1$ \\
 P3   & 857$\pm 1$   & 58$\pm 2$  & 859$\pm 1$ & 71$\pm 2$ &   837$\pm 5$  & 287$\pm 14$   & 0.9$\pm 0.1$  &    & 891$\pm 1$  & 59$\pm 2$ &   895$\pm 1$  &  71$\pm 2$ & 829$\pm 14$ &   284$\pm 28$& 2.0$\pm 0.2$ \\  
\cutinhead{Br $\alpha$} 
 P1  &   867$\pm 1$    & 85$\pm 4$       & 869$\pm 1$&  85$\pm 2$      & 847$\pm 11$     & 37$\pm 39$     & 1.5$\pm 0.2$      & & 872$\pm 1$    & 64$\pm 10$      &  \nodata  & \nodata  &\nodata  & \nodata & \nodata  \\           
 P2  &   862$\pm 1$     & 80$\pm 10$       & \nodata & \nodata & \nodata  & \nodata  &\nodata  & & 895$\pm 1$      & 78$\pm 9$         &  \nodata  & \nodata  & \nodata &\nodata  & \nodata \\    
 P3  &   859$\pm 1$      & 59$\pm 8$      & \nodata & \nodata & \nodata  & \nodata  & \nodata & & 899$\pm 1$     & 72$\pm 8$      &  \nodata  & \nodata & \nodata &\nodata  &\nodata  \\        
 P4  &   866$\pm 1$     & 72$\pm 5$       & 866$\pm 1$       & 76$\pm 2$      & 835$\pm 14$        & 528$\pm 62$       &1.0$\pm 0.1$     & & 884$\pm 1$    & 65$\pm 4$         & \nodata   & \nodata & \nodata & \nodata  &\nodata   \\     
 P5  &  863$\pm 1$     & 74$\pm 10$       & \nodata & \nodata & \nodata & \nodata   &\nodata  & &  901$\pm 1$      & 69$\pm 6$      &   \nodata & \nodata & \nodata & \nodata & \nodata  \\           
P6  &   859$\pm 1$     & 64$\pm 6$       & 860$\pm 1$    & 63$\pm 2$        & 805$\pm 15$     & 432$\pm 53$       & 1.2$\pm 0.2$      & & \nodata  &  \nodata & \nodata   & \nodata  &\nodata   & \nodata  & \nodata \\         
P8  & \nodata  & \nodata & \nodata & \nodata  &\nodata & \nodata   &\nodata  &                                     &  890$\pm 1$     & 74$\pm 7$        &   \nodata & \nodata & \nodata & \nodata & \nodata  \\           
\enddata
\tablecomments{Velocities are heliocentric.   Values are given for the total emission line, as well as the narrow and broad Gaussian components.  
}   
\label{kinematics}
\end{deluxetable}

\clearpage

\begin{deluxetable}{ccc}
\tabletypesize{\footnotesize}
\tablecaption{UV continuum measurements}
\tablecolumns{3}
\tablewidth{0pc}
\tablehead{ \colhead{$N_{Lyc}$ (10$^{52}$ s$^{-1}$) } & \colhead{Authors} & \colhead{Technique} }
\startdata
$\ga$ 2.5 & \cite{Mohan} & radio recombination line models \\
5.7 & \cite{Kawara89} & Br$\alpha$, no extinction correction\\
6.9 & \cite{Kawara89} & Br$\alpha$, extinction corrected\\
5.1 & \cite{JK03} & 7 mm continuum flux \\
4.4 & this paper & 1.3 cm continuum flux \\
11.8 & \cite{Beck01} & Dust heating \\
3.5 & \cite{KJ99} & 2 cm continuum flux, compact cores only  \\
7.3 & \cite{Beck01} & 2 cm continuum flux, compact cores plus extended emission      \\  
\enddata
\label{nlyct}
\end{deluxetable}

\end{document}